\documentclass[twocolumn,superscriptaddress,showpacs,aps,prl]{revtex4}
\usepackage{makeidx}
\usepackage{bm}
\usepackage{graphics}
\usepackage[dvips]{epsfig}
\newcounter{fig}

\begin{document}
\title{ Berry phase, semiclassical quantization  and Landau levels}
\author{A.Yu. Ozerin} 
\affiliation{Verechagin Institute of the High Pressure
Physics, Troitsk 142190, Russia}
\author{ L.A. Falkovsky }
\affiliation{Verechagin Institute of the High Pressure
Physics, Troitsk 142190, Russia}
\affiliation{Landau Institute for Theoretical Physics, Moscow
119334, Russia} 
\begin{abstract}
We propose the semiclassical quantization for  complicated electron systems governed by a many-band Hamiltonian. An explicit analytical
expression of the corresponding Berry phase is derived. This impact allows us to evaluate the Landau magnetic levels  when  the rigorous quantization fails, for instance, for  bilayer graphene and graphite with the trigonal warping.  We find that the magnetic breakdown   can be observed for the certain type of  classical electron orbits.
\end{abstract}
\maketitle
\begin{figure}[b]
\resizebox{.5\textwidth}{!}{\includegraphics{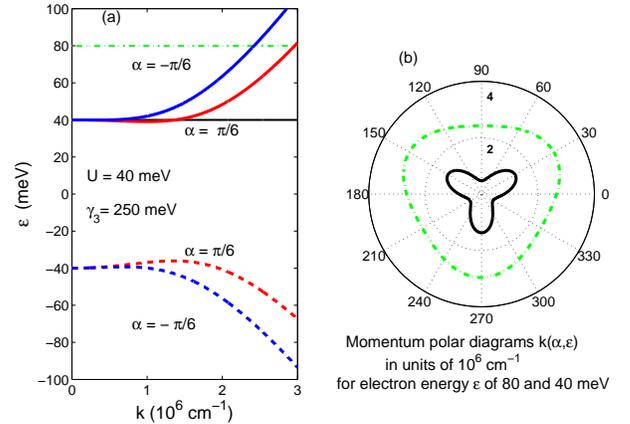}}
\caption{(Color online) (a) The energy dispersion $\varepsilon(k,\alpha)$ of two nearest bands (the electron band shown in solid line and the hole band in dashed line) in bilayer graphene for two polar angles $\alpha$ with the local extrema  at $k\neq 0$ ("mexican hat")  represented. The band  parameters are given in the figure, others are $\gamma_0= 3.05$ eV, $\gamma_1=360$ meV, $ \gamma_4=-150$ meV \cite{PP,GAW}. (b) Cross-sections $k(\alpha,\varepsilon)$ of the electron band for energies of 80 meV (dashed-dotted line) and 40 meV (solid line).} \label{cr_se}
\end{figure}

 
 The most accurate  investigation of the band structure of metals and semiconductors is  studying   the Landau levels  in magneto-transport and magneto-optical experiments.  
 However, the theoretical  solution of the band problem in magnetic fields cannot  often be  exactly found.  A typical example is presented by  graphene layers. For bilayer graphene and graphite,
 the effective Hamiltonian is a $4\times4$ matrix giving four energy bands.
  Fig. \ref{cr_se}  shows nearest two bands of the level structure together  with  semiclassical orbits. 
 The trigonal warping described by the effective Hamiltonian with a relatively small parameter $\gamma_3$ provides an evident effect (see  right panel). Another important parameter is the gate-tunable bandgap $U$ in bilayer graphene. 
  In this situation, the quantization problem cannot be  solved within a rigorous method. 
  To overcome this difficulty one can use a perturbation theory, however this theory becomes quite complicated for the many-band Hamiltonian. 

Alternatively,  the semiclassical quantization can be applied.
 Thus, we can use the Bohr-Zommerfeld condition as
\begin{equation}
\frac{c}{e\hbar B}S(\varepsilon)= 2\pi\left[n+\frac{\mathcal T}{4}+\delta(\varepsilon)\right]\,.
\label{on}\end{equation}
Here $S(\varepsilon)$ is the cross-section area of the electron orbit in the ${\bf
k}$ space for the energy $\varepsilon$ in absence of the magnetic field {\bf B} and for the constant momentum projection $k_z$   on the magnetic field, $n$ is an integer supposed to be large. $\mathcal T$ is the number of the smooth turning points on the electron orbit. There are two smooth turning  points for the Landau levels and only one for  skipping electrons reflected by the hard edge. 

The goal of this letter is an explicit analytical expression for the $\delta(\varepsilon)-$phase within the band
scheme of the matrix Hamiltonian.  
The semiclassical approach is used for the  magnetic field normal to the layered system when the quantization of
 in-layer  momentum components is only essential and  the size of the Fermi surface is small  compared with the Brillouin zone size. We illustrate our results for bilayer graphene.
Notice, that the  $\delta(\varepsilon)-$phase depends on the energy
and can be taken in the
interval $0\leq|\delta|\leq1/2$. If  the spin is neglected,  $\delta=0$ and $\mathcal T=2$ for the Landau levels, and $\delta=1/2$ and $\mathcal T=2$ for
monolayer graphene. In these two cases, the semiclassical  result
coincides  with the rigorous quantization and it is closely connected
with the topological Berry phase \cite{Be}.
 This $\delta-$phase was evaluated for bismuth in  Ref. \cite{Fal}, preceding Berry's work by almost two decades, and it was considered again for bismuth in Ref. \cite{MS}. For graphite,  the semiclassical quantization was applied in Ref. \cite{Dr}. However, in the general case, the evaluation of the $\delta-$phase is still attracted a widespread interest \cite{TA,CU,KEM,PM,PS,LBM,ZFA}.


 The problem under consideration is described  by the
Hamiltonian  in the band representation
\begin{equation}\label{hami}
{(\bf V\cdot\tilde{k}} +\Gamma-\varepsilon)\Psi=0\,, \end{equation}
where the column $\Psi$ consists of   functions  corresponding  with a number of  bands included and is labelled by the band subscript
which we omit together with the matrix subscripts on   $\Gamma$ and ${\bf V}$; a summation over them is implied in Eq. (\ref{hami}). Matrices
 $\Gamma$ and ${\bf V}$ are the first two terms in a series expansion of the Hamiltonian in the power of quasi-momentum $\bf{k}$. 
 
 In the magnetic field, the momentum operator ${\bf\tilde{k}}$
depends on the vector-potential ${\bf A}$  by means the Peierls substitution, 
$${\bf\tilde{k}}=-i\hbar\nabla-e{\bf A}/c,$$ providing the gauge invariance of the theory. 
The magnetic field can also enter 
  explicitly describing the magnetic interaction with  the spin of
a particle. However, for the graphene family, the magnetic interaction is weak and  omitted here.

A simple example of Eq. (\ref{hami}) is given by  the graphene
monolayer. There are two sublattices in it,   and Eq.
(\ref{hami}) is represented by a $2\times2$  matrix  if the spin of carriers is neglected. Another example
considered below is  bilayer graphene with the $4\times4$ matrix Hamiltonian.  For the
monolayer and bilayer graphene, both ${\bf V}$ and ${\bf {\tilde
k}}$ are two-dimensional vectors, e.g., with  $x$ and $y$
components.

We seek for $\Psi$ in the form
\begin{equation}\nonumber
\Psi=\Phi \exp{(is/\hbar)}\,,
\label{be}\end{equation}
where the function $s$ is assumed to be common for all components
of the column $\Psi$. The equation  for $\Phi$ is reduced to
\begin{eqnarray} \label{ham1}[{\bf V\cdot (k}-i\hbar\nabla)
+\Gamma-\varepsilon]\Phi=0\,,\\ \label{k} \text {with} \quad {\bf k}=\nabla s-e{\bf
A}/c\,.
\end{eqnarray}
The function $\Phi$ is expanded in series of $\hbar/i$:
\[
\Phi=\sum_{m=0}^{\infty}\left(\frac{\hbar}{i}\right)^m
\varphi_m\,.
\]
Comparing the terms involving the same powers
of $\hbar$ in Eq. (\ref{ham1}) we have
\begin{equation} \label{ham3}({\bf V \cdot k}
+\Gamma-\varepsilon)\varphi_m=-{\bf V \nabla}\varphi_{m-1}\,.
\end{equation}
For $m=0$, we get a homogeneous system of algebraic equations
\begin{equation} \label{ham4}({\bf V\cdot k}
+\Gamma-\varepsilon)\varphi_0=0\,
\end{equation}
which has a solution under the condition
\begin{equation} \text{Det}({\bf V\cdot k}
+\Gamma-\varepsilon)=0\,.
\label{edm}\end{equation}

This equation determines  the classical electron orbit,
$\varepsilon(k_x,k_y)=\varepsilon$,  in presence of the magnetic field while the  electron energy  $\varepsilon$ is constant.
At the same time, the equation yields   
 the electron dispersion  equation with ${\bf k}$ as the momentum without any magnetic field. In 3d
case, the electron dispersion depends as well on the momentum
projection $k_z$ on the magnetic field and our scheme can be implied in this case without  the expansion in $k_z$.

It is convenient to choose the
vector-potential in the Landau gauge  $A_x=-By , A_y=A_z=0$ 
in such a way that the Hamiltonian does not depend  on the $x-$coordinate. 
Then, the $x-$momentum component $K_x$ becomes a conserved quantum number and the function $s$ in Eq. (\ref{k}) can be written as
\begin{equation} s=xK_x+\sigma(y)\,.\label{s}\end{equation}
The equations (\ref{k}) are reduced to
\[k_x=K_x+\frac{e}{c}By\,,\quad k_y=\frac{d\sigma}{dy}.\]
These equations enable us to use the variable $k_x$ instead of
$y$ and to obtain 
\begin{eqnarray}
\sigma(k_x)=\frac{c}{eB}\int\limits^{ k_x} k_y(k_x') dk_x'\,,
\label{si}\end{eqnarray}
where $k_y$ as a function of $k_x$ is determined by the dispersion equation (\ref{edm}).
 
 The eigenfunction column obeying  Eq. (\ref{ham4})  can be multiplied by the scalar function $C$ common for all elements of the column
\[ \varphi_0\rightarrow C\varphi_0\,\]
where  $\varphi_0$ is any eigen-column of  Eq. (\ref{ham4}). 
The function $C$ is determined by Eq. (\ref{ham3}) with $m=1$. Left-to-right multiplying both sides of
this equation  by $\varphi^*_0$ and using the  Hamiltonian  hermiticity, i.e. the complex conjugations of Eq. (\ref{ham4}), we get  the  consistency condition 
\begin{equation}\label{ham5}
\varphi^*_0{\bf V\cdot\nabla}(C\varphi_0)=0\,,
\end{equation}
where the derivative with respect to $y$ (i.e. to $k_x$) is only to be taken.
The left hand-side of this equation can be written as
\[
\frac{1}{C}\frac{d C}{dk_x}+\frac{1}{2\varphi^*_0V_y\varphi_0} \frac{d\varphi^*V_y\varphi}{dk_x}+\frac{i}{\varphi^*_0V_y\varphi_0}\text{Im}\, \varphi^*_0V_y\frac{d\varphi_0}{dk_x} 
\]
Using the identity \(\varphi^*_0{\bf
V}\varphi_0=\varphi^*_0\varphi_0{\bf v}\) with the electron velocity ${\bf v}=\partial \varepsilon/\partial {\bf k}$, one can write the solution of Eq. (\ref{ham5}) as
\begin{equation}\label{c}
C=c_0(\varphi^*_0\varphi_0 v_y)^{-1/2}\exp(-i\theta)\,,\end{equation}
where \begin{equation}
\theta=\text {Im}\int\frac{dk_x}{\varphi^*_0\varphi_0 v_y}
\varphi^*_0V_y\frac{d\varphi_0}{dk_x}
\label{thet}\end{equation}
and $c_0$ is the normalization factor.  

The quantization condition can be written as usual from the requirement that the wave function has to be single-valued. 
 Continuing   Eqs. (\ref{si}), (\ref{c}), and (\ref{thet}) along the orbit and making the bypass in the complex plane around the turning points where $v_y=0$
  to obtain the decreasing solutions in the classically unaccessible region, one obtains   $\mathcal{T}=2$ and $\delta-$phase  as a contour integral along the classical orbit 
\begin{equation} \label{on1}
\delta(\varepsilon)=\frac{1}{2\pi}\text{Im}
\oint\frac{dk_x}{\varphi^*_0\varphi_0 v_y}
\varphi^*_0V_y\frac{d\varphi_0}{dk_x}\,.
\end{equation}
 Using the Hamiltonian hermiticity, after the simple
algebra (see Ref. \cite{Fal}), Eq. (\ref{on1}) can be rewrite as
\[\delta(\varepsilon)=\frac{1}{4\pi}\text{Im}
\oint\frac{dk}{\varphi_0^*\varphi_0 v} \varphi^*_0\left[{\bf
V}\times\frac{d}{d{\bf k}}\right]_z\varphi_0\]
called usually the Berry phase.

Now let us calculate the $\delta-$phase for bilayer graphene. In simplest  case, the effective Hamiltonian can be written  (see, for instance Refs. \cite{PP,GAW}) as 
\begin{equation}
H(\mathbf{k})=\left(
\begin{array}{cccc}
U \,    & q_{+} \,& \gamma_1    \, & 0\\
q_{-} \,& U     \, & 0\,& 0\\
\gamma_1    \,  &0 \,&-U  \,  &q_{-}\\
0 \,& 0 \,&q_{+} \,&-U
\end{array}%
\right) ,  \label{ham}
\end{equation}%
where the parameter $U$ describes the tunable gap, $\gamma_1$ is the nearest-neighbor hopping integral energy, the matrix elements are expanded in the momentum
$k_{\pm}=\mp ik_x-k_y$ near the $K$ points of the Brillouin zone, and the constant
velocity parameter $v$ is incorporated in the notation $q_{\pm}=vk_{\pm}$.

Here, the orbit is the circle defined by Eq. (\ref{edm}), written in the following form
\begin{equation}
[(U+\varepsilon)^2-q^2][(U-\varepsilon)^2-q^2]-\gamma_1^2(\varepsilon^2-U^2)=0\,.
\label{disp}\end{equation}
The eigenfunction ${\mathbf \varphi_0}$ of the Hamiltonian
(\ref{ham}) can be taken as
\begin{equation}
{\mathbf \varphi_0}=\left(
\begin{array}{c}
(U-\varepsilon)[(\varepsilon+U)^2-q^2]\\
q_{-}[q^2-(\varepsilon+U)^2]\\
\gamma_1(U^2-\varepsilon^2)\\
\gamma_1q_{+}(U-\varepsilon)\end{array}\right) ,\label{fun}
\end{equation}
with the norm squared
\begin{eqnarray}
\varphi_0^*\varphi_0=[(\varepsilon+U)^2-q^2]^2[(\varepsilon-U)^2+q^2]\nonumber\\
+\gamma_1^2(\varepsilon-U)^2
[(\varepsilon+U)^2+q^2]\,.\label{norm}\end{eqnarray}
The derivatives for Eq. (\ref{on1}) are
calculated along the trajectory where the energy  $\varepsilon$ and consequently the trajectory radius $q$ are constant.
The equation (\ref{disp} ) has only one solution for $q^2$ if 
$|U|<|\varepsilon|<\sqrt{U^2+\gamma_1^2}.$ First, let us consider  this case.

\begin{figure}[b]
\resizebox{.5\textwidth}{!}{\includegraphics{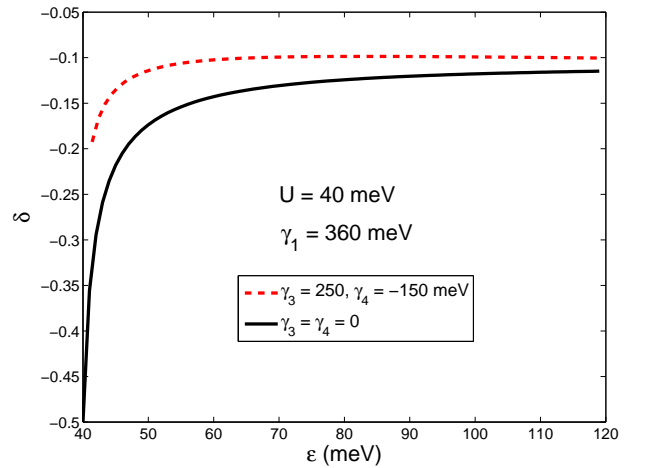}}
\caption{(Color online) Semiclassical phase vs  energy in the conduction band of bilayer graphene without trigonal warping (solid line) and with warping (dashed line).} \label{disg}
\end{figure} 

(i) {\it there is  only one orbit  at given energy  $\varepsilon$ with the radius squared}
$$q^2=U^2+\varepsilon^2+\sqrt{4U^2\varepsilon^2+(\varepsilon^2-U^2)\gamma_1^2}\,.$$ 

The matrix $V_y=\partial H/\partial k_y$ in Eq. (\ref{on1}) has only four nonzero elements $V_{y12}=V_{y21}=V_{y34}=V_{y43}=-1$. Using Eqs. (\ref{disp}) and (\ref{fun}), we find
\begin{equation}\text{Im}\,\varphi_0^*V_y\frac{d\varphi_0}{dk_x}=
4U\varepsilon(U-\varepsilon)[(\varepsilon+U)^2-q^2]\,.\label{sver}\end{equation}
This
expression is  constant on the trajectory as well as $\varphi_0^*\varphi_0$, Eq. (\ref{norm}).
Therefore, in order to find $\delta$, Eq. (\ref{on1}), we have to
integrate along the trajectory
\[\oint
\frac{dk_x}{v_y}\,.\] This integral equals
$-dS(\varepsilon)/d\varepsilon$, where $S(\varepsilon)=\pi q^2$ is
the  cross-section area, Eq. (\ref{on}), with
\begin{equation}\nonumber
\frac{dS(\varepsilon)}{d\varepsilon}=\pi\varepsilon\frac{2(q^2+U^2-\varepsilon^2)+\gamma_1^2}{q^2-U^2-\varepsilon^2}\,.
\nonumber\label{dsec}\end{equation}

Now we have to substitute Eqs. (\ref{norm}), (\ref{sver}),
and (\ref{dsec}) into Eq. (\ref{on1}).
Thus, we find  the Berry phase 
\begin{equation}
\delta(\varepsilon)=\frac{-\varepsilon U}{q^2-\varepsilon^2-U^2}=
\frac{-\varepsilon U}{\sqrt{4U^2\varepsilon^2+(\varepsilon^2-U^2)\gamma_1^2}}
\label{del1}\end{equation}
 shown in Fig. \ref{disg}, where $\delta-$phase of bilayer graphene with trigonal warping is also shown, the detailed calculations will be elsewhere  published.
\begin{figure}[b]
\resizebox{.5\textwidth}{!}{\includegraphics{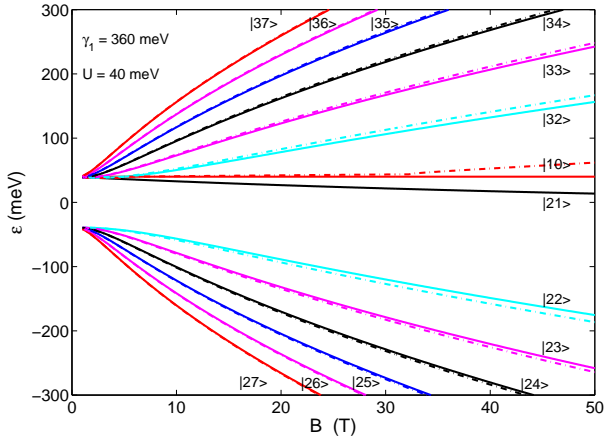}}
\caption{(Color online) Energy levels $\varepsilon_{sn_L}$ for the $K$ valley in magnetic fields for bilayer graphene within rigorous quantization (solid lines) and in the semiclassical approximation (dashed-dotted lines); in the notation $|sn_L\rangle$, $n_L$ is the Landau number  and $s=1,2,3,4$ is the band number, only two nearest bands ($s=2,3$) are shown  at given $n_L$ from 2 to 7. There is only one level, $|10\rangle$, with $n_L=0$ and three levels ($s=1,2,3$) with $n_L=1$. The levels for the $K'$ valley are obtained by mirror reflection with respect to the $\varepsilon=0$ axis.} \label{comp}
\end{figure}
For the ungaped bilayer, $U=0$, the Berry phase $\delta(\varepsilon)=0$. The Berry phase
depends on the energy and $\delta=\mp 1/2$ at $\varepsilon =\pm U$. At the larger energy, $\varepsilon\gg U$, the Berry phase $\delta\rightarrow \mp U/\gamma_1$. 

Substituting Eq. (\ref{del1}) in the semiclassical quantization condition, Eq. (\ref{on}), and solving the equation obtained for $\varepsilon$, we get energy levels as a function of the magnetic field. We have to notice that the Landau numbers $n_L$ listed in Fig. \ref{comp} do not coincide with the numbers $n$ in the semiclassical condition (\ref{on}). The rigorous quantization shows that there are only one Landau level with $n_L=0$ and three Landau levels with $n_L=1$ \cite{Fa}. These levels are not correctly described within the semiclassical approach. However, for $n_L\geq 2$, there are levels in all four bands $s$ (two nearest bands with $s=2,3$ are shown in Fig. \ref{comp}). They correspond with the quantum number $n = n_L - 1$, and  the semiclassical levels become in excellent agreement with the rigorous solution for the larger $n$.

(ii) {\it   for}
$|U|/\sqrt{1+(2U/\gamma_1)^2}<|\varepsilon|<|U|\,,$
{\it at the given energy, there are two orbits with the radius squared}
$$q_{1,2}^2=U^2+\varepsilon^2\pm r\,, \text{where}\quad  r=\sqrt{4U^2\varepsilon^2+(\varepsilon^2-U^2)\gamma_1^2}\,.$$ 
This is an effect of\, "the mexican hat". Then we  seek for the general solution as a sum of two solutions $\varphi_0^j(1)$ and  $\varphi_0^j(2)$ corresponding to these two contours,
\[ \varphi_0^j=C_1\varphi_0^j(1)+C_2\varphi_0^j(2)\]
with two  scalars $C_1$ and $C_2.$ Instead of Eq. (\ref{ham5}) we have a system of two equations written in the $2\times2$ matrix form as follows
\begin{equation}
a\frac{dC}{dq_x}+bC=0
\label{c12}\end{equation}
where the notations of the matrix elements are introduced
\[a_{ik}=\varphi_0^*(i)V_y\varphi_0(k)\,, \quad{\displaystyle
b_{ik}=\varphi_0^*(i)V_y\frac{d\varphi_0(k)}{dq_x}}\,.\]
The off-diagonal matrix elements $a_{ik}$ vanish, $a_{12}=a_{21}=0$. 
Thus, the first equation of the system (\ref{c12}) becomes
\[2q_{1y}r\frac{dC_1}{dq_x}+(2i\varepsilon U-rq_x/q_{1y})C_1 +i(2\varepsilon U+r)C_2=0\,,\]
and the second equation can be obtained with the index replacement $1\leftrightarrow2$ and $r\rightarrow -r$\,.

These equations 
can be simplified with the substitution 
\begin{equation}C_i= \tilde{C}_i(q_i^2-q_x^2)^{-1/4}\,.\label{pm}\end{equation}
For the new functions $\tilde{C}_i$, we get the equation system
\[
\begin{array}{c}{\displaystyle
q_{1y}\frac{d\tilde{C}_1}{dq_x}+iE\tilde{C}_1 +i\sqrt{\frac{q_{1y}}{q_{2y}}}(E+\frac{1}{2})\tilde{C}_2=0\,,}\\{\displaystyle
q_{2y}\frac{d\tilde{C}_2}{dq_x}-iE\tilde{C}_2 -i\sqrt{\frac{q_{2y}}{q_{1y}}}(E-\frac{1}{2})\tilde{C}_1=0\,},
\end{array}
\]
where  the parameter $q_{iy}=\sqrt{q_{i}^2-q_x^2}, \quad i=1,2$  and  $E=\varepsilon U/r$\,.

For the minimum of  conduction band  (maximum of valence band), where $r\rightarrow 0$,
there is a simple limit,
\[
q_{1y}\frac{d\tilde{C}_1}{dq_x}-\frac {i}{2}\tilde{C}_1=0\quad \text {with}\quad \tilde{C}_2=-\tilde{C}_1\,. 
\]
Solving this equation, one gets
\begin{equation}\tilde{C}_1=c_0\exp\left(\frac{i}{2}\arcsin{\frac{q_x}{q_1}}\right)\,.\label{tc}\end{equation}
  Going with $q_x$ along the trajectories and making the bypass in the complex plane around the turning points
$q_x=\pm q_1$ and $q_x=\pm q_2$, we see that 
 both $C_1$ and $C_2$ acquire  from two turning points in Eq. (\ref{pm}) the additional phase $-\pi$ with $\mathcal {T}=2$. At the same time, we have $-1/2$ from   Eq. (\ref{tc}) for $\delta-$phase. 
 Thus, at the boundaries of the narrow interval considered, the $\delta-$phase
 obtains the same value, $\delta=-1/2.$

Taking into account the phases of the functions $\varphi_0^j(i)$, we see, that  the area $S(\varepsilon)$ in Eq. (\ref{on}) can play the different role.
In weak  magnetic fields,  slower oscillations with the smaller $S(\varepsilon)$ corresponding to $q_2$ should
be observed in oscillating phenomena.  However, when the  magnetic field becomes larger
and the semiclassical condition is fulfilled only for the larger cross-section  $S(\varepsilon)$, calculated with   $q_1$, the larger frequency oscillations should be observed. This is nothing but the magnetic breakdown \cite{CF} which should be utilized if the chemical potential belongs to the interval  where the effect of  "the mexican hat" appears.


In conclusion, our study  shows that the semiclassical approach gives a powerful tool for probing the electron magnetic properties in metals. The Berry phase depending on the energy can be calculated and observed even for complicated band scheme.
The method presented here should be useful  for many electron systems.

We thank I. Luk'yanchuk for helpful discussions. This work was supported  by the SCOPES grant IZ73Z0$\_$128026 of Swiss NSF, by the grant SIMTECH No. 246937, and by the Russian Foundation for Basic
Research (grant No. 10-02-00193-a).


\begin{thebibliography}{99}
\bibitem{Be} M.V. Berry, Proc. Roy. Soc. London, Ser. A \textbf{392}, 45 (1984)
\bibitem{Fal} L.A. Falkovsky, Zh. Eksp. Teor. Fiz. \textbf{49}, 609 (1965) [Sov. Phys. JETP {\bf 22}, 423 (1966)].
\bibitem{MS} G.P. Mikitik, Yu.V. Sharlai, Zh. Eksp. Teor. Fiz. \textbf{114}, 1357 (1998)[Sov. Phys. JETP \textbf{87}, 747 (1998)];
 Phys. Rev. B \textbf{67}, 115114 (2003).
\bibitem{Dr} G. Dresselhaus, Phys. Rev. B \textbf{10}, 3602 (1974).
\bibitem{CU} P. Carmier, D. Ullmo, Phys. Rev. B \textbf{77}, 245413 (2008).
\bibitem{TA} A.A. Taskin, Y. Ando, Phys. Rev. B \textbf{84}, 035301 (20011).
\bibitem{KEM} E.V. Kurganova, H.J. van Eleferen, A. McCollam, L.A. Ponomarenko, K.S. Novoselov, A. Veligura, B.J. van Wees, J.C. Maan, U. Zeitler, Phys. Rev. B \textbf{84}, 121407 (20011). 
\bibitem{PM} Cheol-Hwan Park, N. Marzari, Phys. Rev. B \textbf{84}, 205440 (2011). 
\bibitem{PS} Singhun Park, H.-S. Sim, Phys. Rev. B \textbf{84}, 235432 (2011).
\bibitem{LBM} Y. Liu, G. Bian, T. Miller, T.-C. Chiang, Phys. Rev. Lett.  \textbf{107}, 166803 (2011).
\bibitem{ZFA} L.M. Zhang, M.M. Fogel, D.P. Arovas,
Phys. Rev. B \textbf{84}, 075451 (2011).
\bibitem{PP} B. Partoens, F.M. Peeters, Phys. Rev. B \textbf{74}, 075404 (2006).
\bibitem{GAW} A. Gr\"{u}neis, C. Attaccalite, L. Wirtz, H. Shiozawa, R. Saito, T. Pichler, A. Rubio, Phys. Rev. B \textbf{78}, 205425 (2008). 
\bibitem{Fa} L.A. Falkovsky
Phys. Rev. B \textbf{84}, 115414 (2011).
\bibitem{CF} M.N. Cohen, L.M. Falikov, 
Phys. Rev. Lett.  \textbf{7}, 231 (1961).

\end{thebibliography}
\end{document}